# Using Convolutional Neural Networks for Denoising and Deblending of Marine Seismic Data

S. Slang[1,2*], J. Sun[1,2], T. Elboth[2], S. McDonald[2] and L. Gelius[1]

[1]University of Oslo, Norway, [2]CGG

## Summary

Processing marine seismic data is computationally demanding and consists of multiple time-consuming steps. Neural network based processing can, in theory, significantly reduce processing time and has the potential to change the way seismic processing is done. In this paper we are using deep convolutional neural networks (CNNs) to remove seismic interference noise and to deblend seismic data. To train such networks, a significant amount of computational memory is needed since a single shot gather consists of more than $10^6$ data samples. Preliminary results are promising both for denoising and deblending. However, we also observed that the results are affected by the signal-to-noise ratio (SnR). Moving to common channel domain is a way of breaking the coherency of the noise while also reducing the input volume size. This makes it easier for the network to distinguish between signal and noise. It also increases the efficiency of the GPU memory usage by enabling better utilization of multi core processing. Deblending in common channel domain with the use of a CNN yields relatively good results and is an improvement compared to shot domain.

**Introduction**
The recent availability of powerful GPUs and open source software have enabled artificial neural networks (ANNs) to be applied to a number of practical and industrial scale problems. The level of adoption of this technology within the field of O&G exploration is well illustrated by the number of abstracts related to ANNs that are submitted to the annual EAGE and SEG conferences. Since 2001, there have typically been one or two papers per year discussing ANNs. In 2018 the level rose significantly to between 50 and 100 papers.

In seismic processing, ANNs have the potential to be applied to many of the key processing steps (swell noise attenuation, seismic interference attenuation, deblending, deghosting etc.) which today involve significant testing time and computational power. Once trained, ANNs are computationally very light and potentially adaptable to varied datasets. Their use could therefore significantly save processing times and, in the long term, impact the whole business sector.

A natural first step in this direction is to look at various forms of seismic data denoising. This is not an entirely new concept, and is clearly inspired by work done on natural picture denoising, where we refer to Zhang et al. (2017) for a recent overview. However, this field is still immature and, as indicated by Xie et al. (2018), a lot of work is needed before ANNs can be effectively applied to seismic data denoising.

A common limitation in recent papers using ANNs for seismic data denoising (see e.g. Ma (2018), Baardman (2018), Si and Yang (2018), Li et al. (2018), Jin et al. (2018), and Zhang et al. (2018)) is that they only use synthetic data or noise on datasets with limited dynamic range and/or frequency content. As proof of concept, this has value. However, we have not yet seen convincing examples that compare against existing state-of-the-art denoising results.

In conventional processing, it is a common practice to sort and/or transform the seismic data into domains wherein it is easier to separate the noise from the desired signal. We have not yet seen this approach utilized in ANN denoising, and we believe that this could potentially improve results significantly.

This paper is structured as follows: in the theory section, we will introduce our network architecture and outline the design approach. We will then present two examples of denoising done on real marine seismic data, before pointing towards how we believe this work could be taken further.

**Theory – CNNs for seismic denoising**
A common type of layer in ANNs is the Fully Connected (FC) layer. They tend to give good results, but they are computationally heavy since they have one parameter for each sample in the input data. Seismic datasets tend to be large (~$10^6$ samples per shot gather), making the use of FC layers a challenging undertaking given the large memory requirements and computational demand. This leads us to another common ANN type, which is the Convolutional Neural Networks (CNNs).

CNNs are neural networks consisting of at least one convolutional layer. Convolutional layers differ from other types of layers in that they employ convolutions over subsets of the data, rather than a general matrix multiplication. According to Goodfellow et al. (2016), this makes CNNs well suited for 2D images where neighboring pixels are connected in a larger pattern. It should therefore be possible to denoise seismic data with localized and 'random' noise either in the shot domain or when sorted to, for example, the channel domain. We assume that CNNs will be able to handle seismic data, given its continuous nature and 2D structure. This will greatly reduce the computational cost and memory requirements compared to FC layers. Although it is much more efficient to use CNNs with respect to computational power, seismic images are large and still push the limits of hardware available today.

Seismic noise varies a lot in amplitude and might be orders of magnitude larger than the underlying signal, making it hard for the network to recreate the underlying signal structure. Given the difficulties raised by the large input volumes and the sometimes low SnR that can occur in recorded seismic data, it is understandable that previous attempts have been made with synthetic data or data subsets with limited size and dynamic range. When using synthetic data, the user has full control over the dataset. In this work, we apply CNNs to real life, full-scale marine seismic gathers and investigate how well this works for two types of commonly encountered seismic noise attenuation problems.

**Example 1: Seismic interference noise attenuation**

The first example investigated is the attenuation of seismic interference (SI) noise. This is dispersive coherent acoustic energy originating from other seismic crews operating nearby. The energy is mostly propagating in the water column, and is typically recorded with amplitudes similar to or larger than that of the seismic reflection signal. As such, it is common practice to try to attenuate this noise early on in the processing flow.

CNNs require both noisy images and clean images, which are regarded as ground truth, in image denoising. The network calculates the error between the denoised image and ground truth to update the weights. Our marine seismic training data consists of 800 records containing almost pure SI-noise recorded from different directions, and 482 (nearly) noise-free seismic shot gathers from the North Sea. It is a straightforward task to blend clean shots with varying SI-noise creating a dataset for the network to train on where ground truth is known.

The network architecture is based on common models used in image analysis with CNNs. It consists of convolutional layers with batch normalization and Rectified Linear Unit (ReLU) activation function. To handle the full seismic range, residual learning is applied, making sure all layers learn from the full frequency range in the original input image. The main difference compared to other attempts is that no downscaling is applied. The image is full-size throughout the network to reduce potential blurring and precision loss. This results in a large model requiring significant computational power and about 12GB GPU memory to train on a single image. The overall training process to achieve this level of denoising involved around $10^4$ shot gathers, and took nearly two weeks on a modern GPU (6 GB x2). However, we note that once the network is trained, the actual denoising of a single shot gather is done in less than a second.

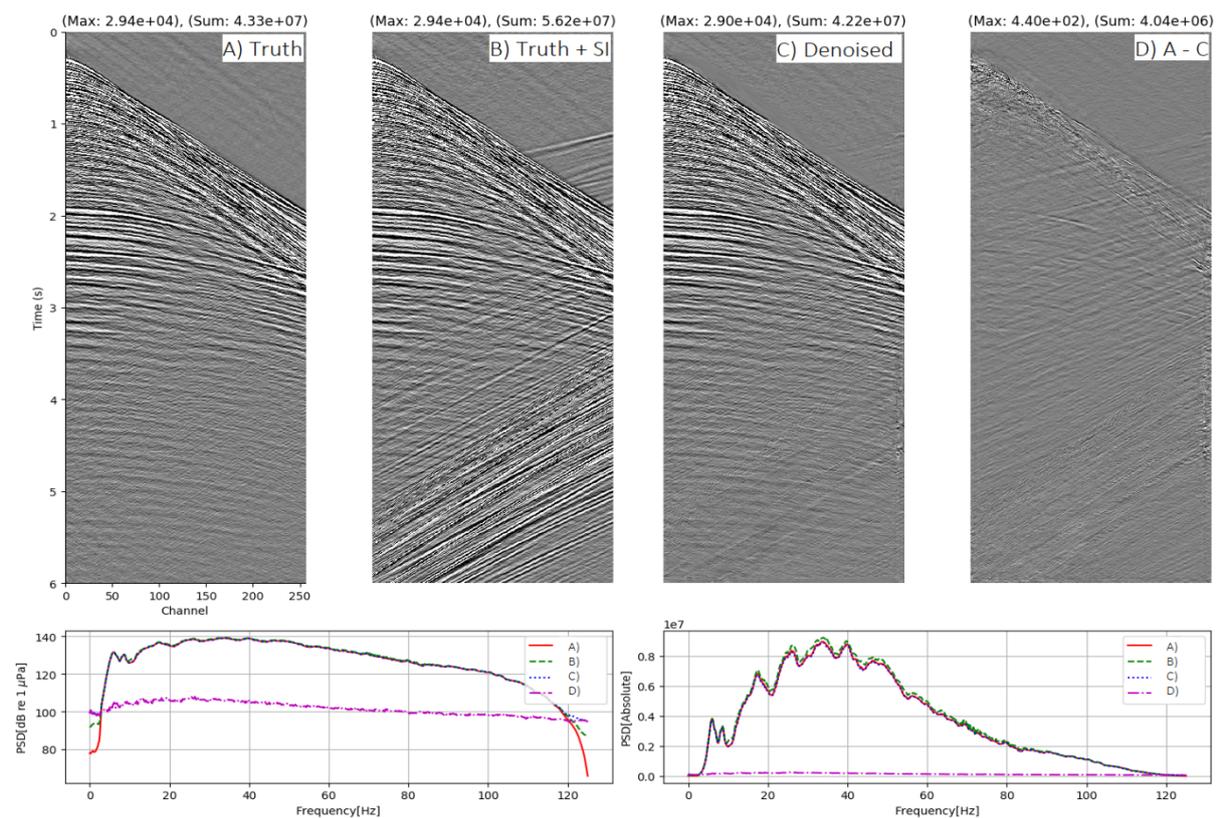

**Figure 1:** Figure illustrating the removal of SI noise from shot gathers where the different images from left to right are: A) clean shot, B) SI noise contaminated shot, C) denoised shot, D) difference between clean and denoised shot. The numbers above each shot are picture values showing the maximum and overall energy of the image. Fourier spectra (logarithmic and absolute) are appended at the bottom of the image to show the frequency content of each image.

Figure 1 illustrates that the network is removing SI rather well for cases where the SnR is relatively high. This is visible in the spectra showing that the network has a good recovery for all frequencies.

There is some leakage apparent in D), but it appears more dominant due to the high gain applied to the images. However, in cases with strong SI with very similar direction to the underlying reflection data, the results are not as good. Moving to common channel domain, where the noise is less coherent, is a way to overcome this problem. This allows us to use smaller images which still contain coherent signal but contain less coherent noise. Smaller images require less computing memory, which opens the possibility of using FC-layers. The downside of such a domain change is that 'real time' denoising during data acquisition is no longer possible since multiple shot records are needed to construct common channel images.

**Example 2: Seismic data deblending**
The second example is the deblending of N+1 shot data to extend the useful record length in the seismic data. Deblending (separation) of overlapping seismic records is important since it is a key technology enabling improved sampling and/or more efficient acquisition. The basic idea was probably first introduced almost 50 years ago by Barbier (1971), but wide scale industrial adoption has only been achieved in the last few years.

Our dataset consists of 1300 towed unblended marine split spread gathers. The shot gathers are artificially blended with the N+1 shot with a fixed delay of 1.2 second and a specially designed dither. Based on our study, we have found that it is difficult for the neural network to learn how to deblend seismic data in the shot domain. The events of two different shots in the blended data share similar characteristics in both amplitudes and dip, thus there is no clear characteristic features for the network to learn. Sorting the data into common channel domain, as mentioned in Example 1, gives better results. The N+1 shot in common channel exhibits randomness, while shot N has a continuous nature. Thus, deblending of overlapping seismic records in common channel domain is similar to removing random noise in image processing. The characteristics of CNNs, as mentioned in the theory section, is suitable for this case. The unblended data serves as ground truth for the network and is used to update the weights of the artificial neurons.

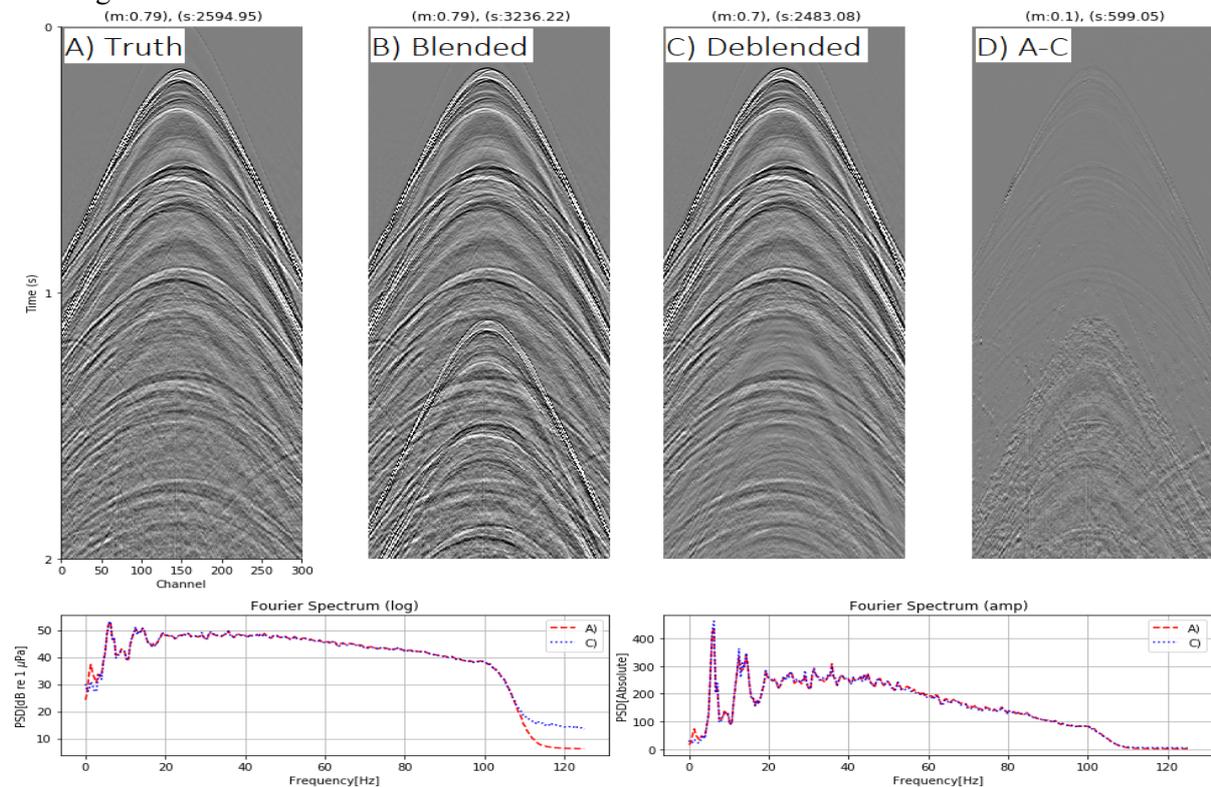

**Figure 2:** Figure illustrating the deblending result from shot gathers where noise amplitude is 80% of signal amplitude. The images from left to right are: A) unblended shot, B) blended shot, C) deblended shot, D) difference between unblended shot and deblended shot. The numbers above each shot are picture values showing the maximum and overall energy. Fourier spectra (logarithmic and absolute) are appended at the bottom of the image to show the frequency content of unblended shot and deblended shot.

Figure 2 illustrates that the network has the ability to learn deblending, and works well when we only introduce 80% of shot N+1, thus artificially enhancing the SnR. The recreated plot visible in C) has a quality that is close to current state-of-the-art commercial processing, but would compare less favorably without artificial SnR enhancement. The training, validation, and test sets consist of 14400, 2700, 900 images respectively with 60 channels per image to reduce memory usage. This particular test took approximately 2.5 days to run, however, once the network was trained, the computational cost of deblending a single image was approximately the same as Example 1 – less than a second. All the processing is done in common channel domain. The records with no overlapping noise are almost perfectly recreated.

The network architecture is similar to Example 1, with a network consisting of convolutional layers with applied batch normalization. The largest difference is the usage of Leaky ReLU activation function to reduce the risk of "dead" neurons. Compared to common CNNs, both our mentioned network models differ in terms of static image size and thus no max pooling. This reduces the risk of blurring and loss of geological data. Even with a robust network, deblending in common channel domain remains a challenging task. This is because the SnR tends to always be low over a large area compared to other types of noise removal. Although this issue may cause some residual noise in the output image, the network is removing a significant amount of the N+1 shot and keeping detailed underlying signal intact. As mentioned in Example 1, there are multiple ways to potentially enhance the result. One way could be preprocessing the data to improve the SnR or moving the processing to a sparser domain. These novel approaches will be the focus of our future work.

**Conclusions**

Convolutional neural networks show promising results for interference noise attenuation and N+1 deblending of marine seismic data. When applied alone, the results are below the level achieved by state-of-the-art commercial processing. However, we have shown that applying machine learning to seismic data processing can still produce encouraging results and is an approach worth exploring further. For the problems shown, sorting the data into common channel domain seems to give better results than shot domain, due to the random nature of the noise in said domain. The higher the signal to noise ratio, the better the results become. We finally mention the importance of having high quality training datasets. The ground truth should ideally be without any noise. This is difficult to achieve when we work with real data, which inevitably will have some noise contamination.

**Acknowledgements**

The authors want to thank CGG MCNV for providing the field data used in this paper.